\documentclass{svproc}
\usepackage{fix-cm}
\usepackage{lmodern}
\usepackage[T1]{fontenc}
\usepackage{amsmath}   % Required for math environments
\usepackage{amssymb}   % Extra math symbols
\usepackage{bm}        % Bold math symbols
\usepackage{mathrsfs}  % Math script fonts
\usepackage{url}
\usepackage{makeidx}
\usepackage{newtxmath}
\usepackage[bottom]{footmisc}
\usepackage{newtxtext} 
\usepackage{hyperref}
\usepackage{lineno,xcolor,graphicx}
\fontsize{5.5pt}{6pt}\selectfont
\usepackage[utf8]{inputenc}
\usepackage[T1]{fontenc}
\usepackage{lmodern}
\usepackage{listings}
\usepackage{amsmath}
\renewcommand{\textcolor}[2]{#2}
\usepackage{changepage} % Dodany pakiet
\begin{document}
\mainmatter              % start of a contribution
\title{Dynamics of pendulum forced by a magnetic excitation with position-dependent phase }
%
%\titlerunning{Hamiltonian Mechanics}  % abbreviated title (for running head)
%                                     also used for the TOC unless
%                                     \toctitle is used
%
\author{\underline{Krystian Polczyński$^*$}[0000-0002-1177-6109]  \and Maksymilian Bednarek  [0000-0002-7669-119X] \and Jan Awrejcewicz  [0000-0003-0387-921X]}
\authorrunning{Krystian Polczyński et al.} % abbreviated author list (for running head)
%
%%%% list of authors for the TOC (use if author list has to be modified)
\tocauthor{Krystian Polczyński, Maksymilian Bednarek, Jan Awrejcewicz}
\institute{Lodz University of Technology, Department of Automation, Biomechanics and Mechatronics, 1/15
	Stefanowskiego Str., 90-924 Łódź, Poland,\\
\email{krystian.polczynski@p.lodz.pl ($^*$corresponding author)} (\underline{presenting author}),\\ WWW home page:
\texttt{https://abm.p.lodz.pl/en/krystian-polczynski}
\email{maksymilian.bednarek@p.lodz.pl}\\
\email{jan.awrejcewicz@p.lodz.pl} }

\maketitle              % typeset the title of the contribution
\begin{abstract}
This study investigates the dynamics of a magnetic pendulum under time-varying magnetic excitation with a position-dependent phase. The system exhibits complex chaotic and regular dynamics, validated through simulations and experiments. The mathematical model, based on a physical setup, includes a magnetic excitation torque with phase dependence on the dynamic variable. Bifurcation analyses confirm the rich multistability of the system, showcasing periodic attractors, period-doubling bifurcations, and chaotic behavior. Experimental validation demonstrates a high agreement between numerical and experimental results, supporting the efficacy of the proposed model. The study sheds light on the system's sensitivity to changes in magnetic interaction, providing insights into controlling resonance energy exchange in coupled magnetic pendulum systems.
\end{abstract}%
\section{Introduction}
Physics encompasses various fundamental branches, with mechanics and electromagnetism standing prominently among them. Traditionally, scientists delved into these domains separately throughout the centuries. However, the relentless march of technological advancement, coupled with the pursuit of expanding scientific frontiers, compelled researchers and engineers to conceive systems that seamlessly blend elements from both realms. In the contemporary technical landscape, these integrated systems find their place in the interdisciplinary realm of mechatronics \cite{Bishop2002,AwrejM}. Electric motors, such as stepper motors \cite{Kepinski2015} or linear motors \cite{Gajek2018}, serve as exemplary instances of such hybrid systems, playing crucial roles as sources of mechanical energy.

This study explores a magnetic pendulum experiencing analogous forces and exhibiting phenomena akin to the previously discussed mechatronic systems. Because of its uncomplicated design, this pendulum proves valuable for conducting simulations and experiments related to the fundamental nonlinear phenomena observed in mechano-electro-magnetic devices.

To begin, we examine systems featuring one-degree-of-freedom magnetic oscillators. Bethenod \cite{Bethenod1938} analytically investigated the sustained and undamped oscillations of a pendulum subjected to a periodically changing magnetic field. In this study, the frequency of the alternating magnetic field surpassed the natural frequency of the pendulum. However, the obtained results did not sufficiently elucidate the mechanism behind the emergence of sustained oscillations. Rocard \cite{Rocard,Landa1996} and Knauss et al. \cite{Knauss} delved into a theoretical analysis of Bethenod's pendulum, employing mathematical power series approximations for its description. Their inquiries primarily focused on the pendulum's linear behavior within small deviation angles. In a different approach, Minorsky \cite{Minorsky} opted to investigate this problem using Mathieu equations with a moving parametric point. The results obtained were qualitatively correct, providing mathematical conditions for the appearance of self-oscillations, periodic oscillations with a stationary amplitude, and unstable stationary oscillations.
Detailed investigations into the self-oscillations of the magnetic pendulum have been conducted by Skubov et al. \cite{Skubov2014}. They extensively examined limit cycles and potential equilibrium positions through asymptotic solutions derived from Lagrange-Maxwell equations. Damgov et al. \cite{DamgovPopov} categorized the motion of the magnetic pendulum into two cases based on the pendulum's distance from the coil. Both numerical and analytical studies revealed that the system manifests discrete/quantized amplitudes of oscillations under the inhomogeneous influence of a periodic force. A similar analytical approach for analyzing the motion of a single magnetic pendulum has been presented in the work by Wijata et al. \cite{Wijata2021}. Continuous and discrete mathematical models of the system were employed to establish conditions for one-sided oscillations and perform bifurcation analysis. Various scenarios of one-sided oscillation were considered, and numerical tests were compared with experiments, showing good correlation. Furthermore, a semi-analytic approximate method, utilizing averaging ideas, was developed for the studied system and successfully validated in \cite{Skurativskyi2022a}.

Nana et al. \cite{Nana2017,Nana2018} conducted a comprehensive examination of typical nonlinear effects within the magnetic pendulum system. The system exhibited amplitude jumps, hysteresis, bistable states, as well as periodic and chaotic dynamics during both experimental and theoretical investigations. 
The examination of a vertically driven magnetic pendulum, influenced by electromagnetic interactions arising from eddy currents induced in a nearby conducting plate, has been detailed in the works of Boeck et al. \cite{Boeck2020a,Boeck2021,Becker2022}. The problem was simplified to the Mathieu equation, and the harmonic balance method was employed to explore the conditions leading to instability in its equilibrium position due to electromagnetic interactions. The paper also presents numerical analysis results, highlighting the emergence of doubly connected regions of harmonic instability, the coexistence of stable periodic orbits, and the occurrence of chaotic motions during subharmonic instability under moderately strong driving.

Recent studies by various researchers have investigated the dynamics of coupled magnetic pendulum systems and other multi-degree-of-freedom oscillators exposed to a magnetic field. In works \cite{Polczynski2019c,Polczynski2020a,PolczynskiSkurativsky}, the nonlinear dynamics of two coupled pendulums were explored, one influenced by a magnetic force and the other moving due to torsion coupling via a flexible element. The mathematical model, incorporating experimental data for the magnetic torque, revealed diverse behaviors, including chaotic, multi-periodic, and quasi-periodic solutions. Bifurcation analysis, Poincar'e sections, Lyapunov exponents, and Fourier spectra confirmed the findings, with numerical and experimental studies showing high agreement.

Pilipchuk et al. \cite{PilipchukPolczynski2022} propose a methodology for controlling resonance energy exchange in a system of two weakly coupled magnetic pendulums interacting with a magnetic field. The study demonstrates that guided magnetic fields can effectively modify mechanical potentials, directing energy flow between the pendulums. Antiphase oscillations show energy transfer from the repelling magnetic field to the attracting one, while inphase oscillations exhibit reversed energy flow. The closed-loop controller, relying on phase shift information estimated through the coherency index, operates at a relatively slow temporal rate compared to the oscillations, highlighting its advantageous control strategy.

This study aims to explore the dynamic characteristics of a highly nonlinear magnetic pendulum system exposed to time-varying magnetic excitation, where the phase and frequency depend on the dynamic variable. The system demonstrates both complex chaotic and regular dynamics, as validated through experimental and simulation approaches. Analyzing the structure of basins of attraction for selected system parameters is also part of our investigation.

\textcolor{red}{The work of \cite{ICNDA2022} presents the operation of an algorithm based on the deep learning framework Long Short-Term Memory. Using this algorithm, they predicted the occurrence of chaotic dynamics in a logistic map system forced quasiperiodically. The magnetic pendulum presented by us in this work could constitute an interesting challenge for the mentioned DL algorithm in order to predict the occurrence of chaotic dynamics and check the correctness of the algorithm.}

Our research was motivated by Krylosova et al.'s work \cite{Krylosova2020}, where they studied a non-autonomous oscillator influenced by external forces with phase and frequency dependencies on the system's dynamic variable. Their findings revealed the emergence of complex chaotic dynamics in the oscillator's behavior due to controlled external force parameters. The exploration of control parameter space uncovered various periodic and chaotic oscillations, showcasing similarities and differences with non-autonomous oscillators featuring periodic potentials.
\begin{figure}[b!]
	\hspace{3cm}	(a)\hspace{5cm} (b)\\
	\includegraphics[scale=0.45]{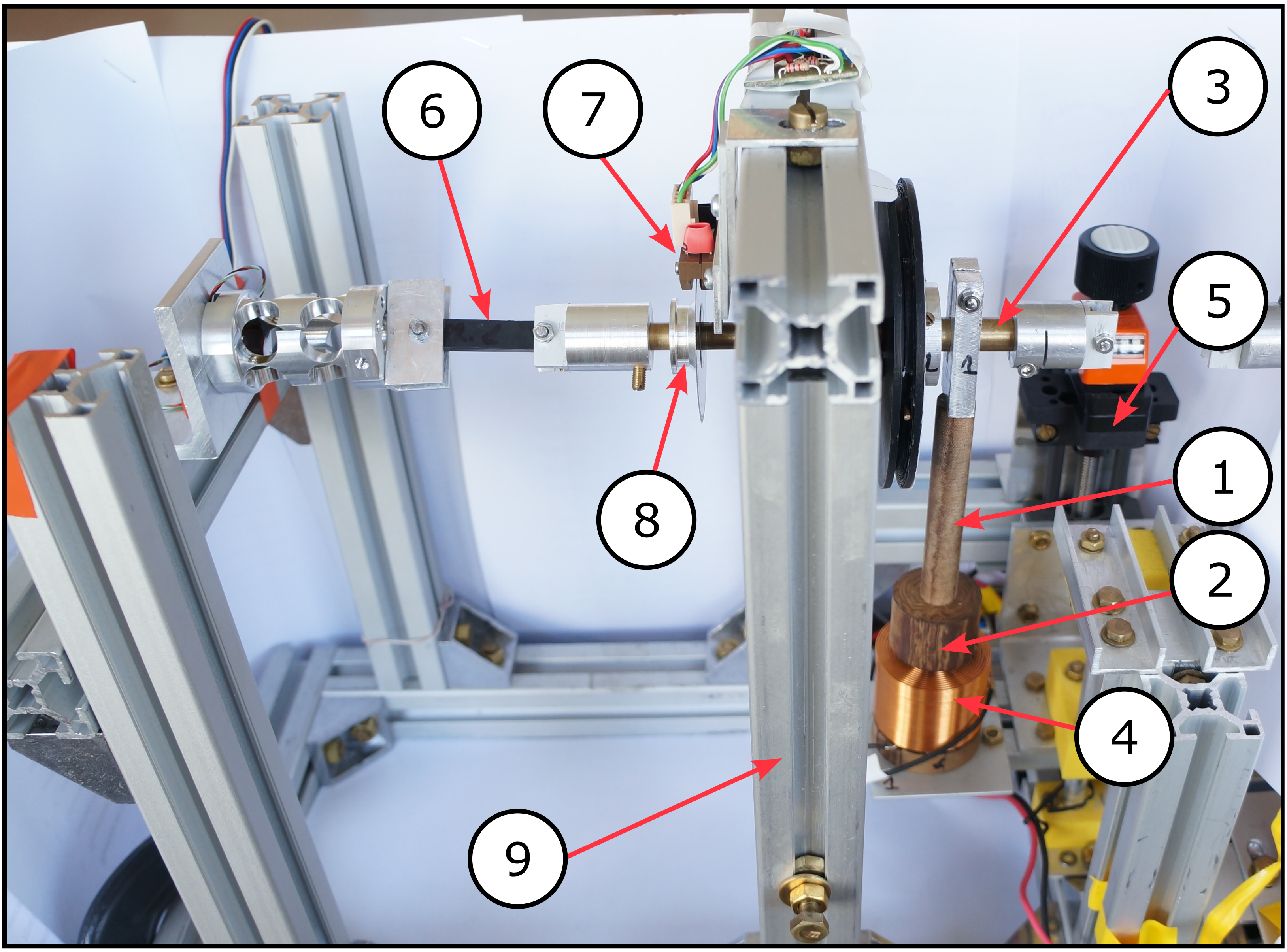}
	\hspace{0.5cm}
	\includegraphics[scale=0.7]{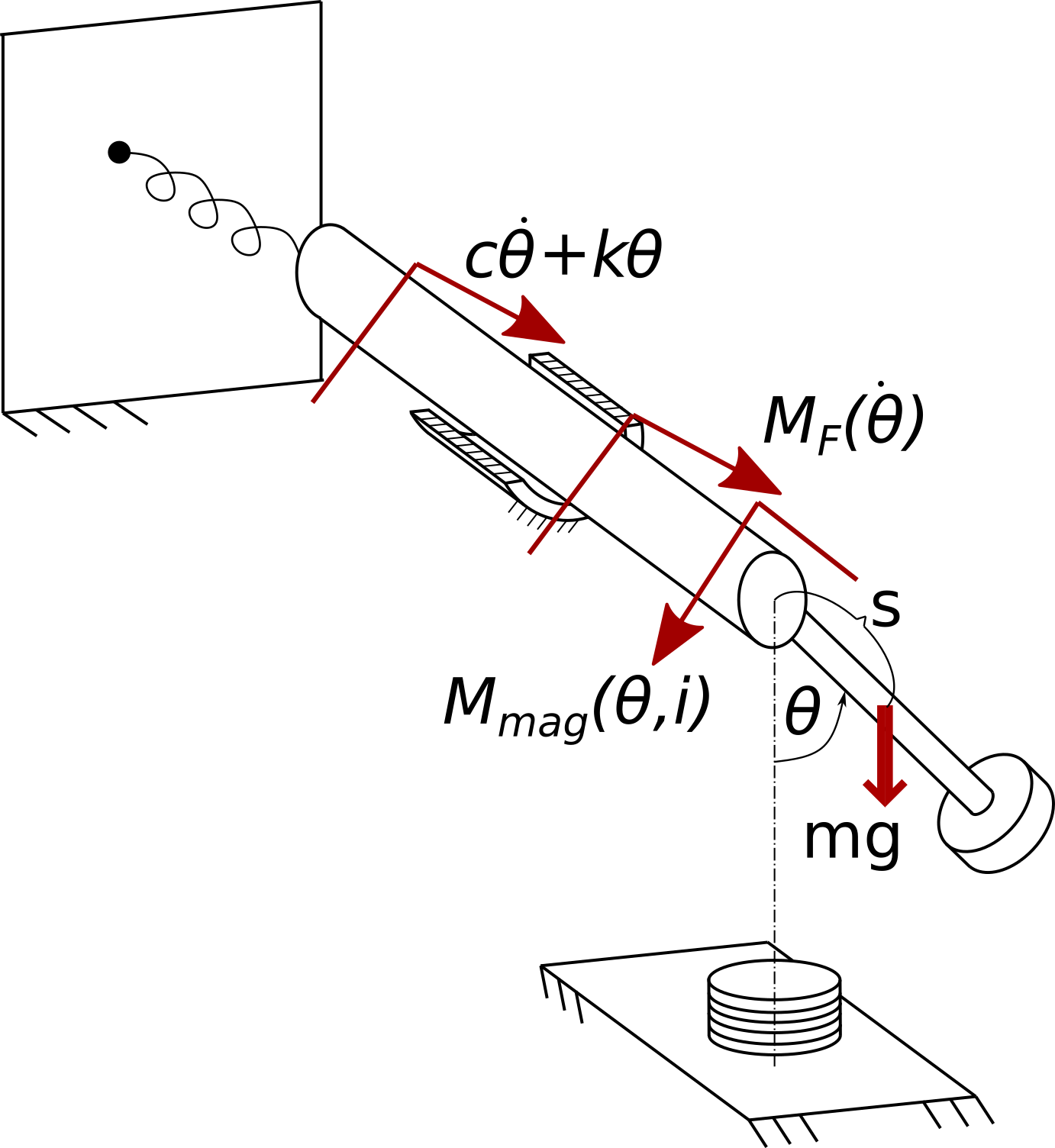}
	\caption{(a) Experimental setup and (b) physical model. Description of the numerical labels can be found in the main text.}
	\label{fig:experimental_rig_1}
\end{figure}
\section{Experimental setup}
An experimental investigation was carried out using a specially constructed setup, as illustrated in Fig.~\ref{fig:experimental_rig_1}a.
The setup consists of a magnetic pendulum (1) with a neodymium magnet (2) attached to one end of a axis (3). An electric coil (4) is positioned on a fixed platform beneath the pendulum. The platform's vertical position can be adjusted by a linear lift (5). The other end of the axis connects to a fixed base via an elastic rubber element (6). The electric coil is powered by a laboratory power supply. The coil current signal follows a voltage signal from an NI USB-6341 card, controlled by LabView software. During experiments, a positive current repels the magnet from the coil, while a negative current causes attraction. The angular position of the pendulum is recorded by an optical incremental sensor (7). The materials used for the setup, including the frame (9), are non-magnetic.

\section{Mathematical modeling}
In this section, we developed a mathematical model for the system based on physical model shown in Fig.~\ref{fig:experimental_rig_1}b.  \textcolor{red}{For this purpose, we used the Newton-Euler method, and the resulting dynamical equation of motion is as follows}
\begin{equation}
	J\ddot{\theta}+c\ddot{\theta}+mgs\sin{\theta} +k\theta+M_F(\dot{\theta})=M_{mag}(\theta,i(t)),
	\label{eq:Newton_motion_1}
\end{equation}
where: $\theta$~-- angular position, $J$~-- moment of inertia, $k$~-- spring stiffness, $c$~-- viscous damping coefficient, $mg$~-- gravitational force, $s$~-- arm of gravitational force.
\textcolor{red}{Based on experimental research conducted in \cite{Wijata2021}, we assumed that the elastic rubber element is characterized by linear stiffness and viscous damping. The nonlinear nature of the rubber was so insignificant that its effects introduced into the system could be ignored. Moreover, the mentioned experimental studies from work \cite{Wijata2021} showed that the rolling resistances in the shaft bearings are so significant, that they had to be taken into account in the equation of motion. These resistances had Coulomb and static nature. The developed resistance model takes into account both Coulomb and static resistances, as well as their transition, i.e. the so-called the Stribeck effect.}
Term $M_F(\dot{\theta})$ represents friction torque of rolling bearings and it is expressed as follows
\begin{equation}
	M_F(\dot{\theta})=\left[\tau_c+(\tau_s-\tau_c)\exp{\left( \frac{-\dot{\theta}^2}{v_s^2} \right)}\right]\tanh{\epsilon\dot{\theta}},
	\label{eq:bearing_damping_2}
\end{equation}
where $\tau_c$, $\tau_s$~-- the Coulomb and static friction torques, respectively. Coefficient $v_s$~-- Stribeck velocity, $\epsilon$~-- regularization parameter. 

The term $M_{mag}(\theta, i(t))$ function is a the magnetic excitation torque in the system and depends on the angular position of the pendulum and the coil current $i(t)$.
\begin{equation}
	M_{mag}(\theta,i(t))=\frac{2a(i(t))}{b(i(t))}\exp{\left[\frac{-\theta^2}{b(i(t))}\right]}\theta.
	\label{eq:magnetic_func_3}
\end{equation}
Parameters $a$ and $b$ determine the amplitude and shape of the magnetic torque and have been experimentally investigated. 
Parameter $b$ is constant, while parameter $a$ depends linearly on the current
\begin{equation}
	a(i(t))=K_p\:i(t),
	\label{eq:a_parameter_function_4}
\end{equation}
where $K_p$ coefficient is constant.

Initially, the current signal $i(t)$ supplied to the system takes the following form
\begin{equation}
	i(t)=I_0\sin{(\omega_0t+\phi_0)},
	\label{eq:coil_current_signal_5}
\end{equation}
where $I_0$~-- maximum amplitude, $\omega_0$~-- angular frequency, $\phi_0$~-- initial phase. Considering Eqs.~(\ref{eq:a_parameter_function_4}) and (\ref{eq:coil_current_signal_5}) and substituting them into Eq.~(\ref{eq:magnetic_func_3}), we obtain the following expression for the magnetic torque
\begin{equation}
	M_{mag}(\theta,t)=\frac{2K_pI_0}{b}\exp\left({\frac{-\theta^2}{b}}\right)\theta\sin{(\omega_0t+\phi_0)}
	\label{eq:magnetic_torque_sine_6}.
\end{equation}

To generalize our study, we opted to transform Eq.~(\ref{eq:Newton_motion_1}) to  a~dimensionless form using the following substitutions
$y={\theta}/{\theta_s}$ $x={t}/{t_s}$,
where $\theta_s=\sqrt{b}$ and $t_s=\sqrt{\frac{J}{mgs}}$ are scaling factors. After transforming into a dimensionless form, the governing equation of motion becomes
\begin{equation}
	y''+\beta y'+\alpha y+\gamma\sin{\left(\frac{1}{\gamma}y\right)}+\left[\delta+\zeta\exp\left({\nu {y'}^2}\right)\right]\tanh{(\sigma y')}=A_0\exp{(-y^2)}y\cdot\sin{(\Omega x+\phi_0)}
	\label{eq:motion_dimensionless_8},	
\end{equation}
where $\alpha=\frac{k}{mgs}$, $\beta=\frac{c\sqrt{J}}{J\sqrt{mgs}}$, $\gamma=\frac{1}{\sqrt{b}}$,\;$\delta=\frac{F_c}{mgs\sqrt{b}}$, $\zeta=\frac{F_s-F_c}{mgs\sqrt{b}}$, $\nu=-\frac{bmgs}{v_s^2J}$, $\sigma=\frac{\epsilon\sqrt{bmgs}}{\sqrt{J}}$, $A_0=\frac{2K_pI_0}{bmgs}$, $\Omega=\frac{\omega_0}{\sqrt{mgs/J}}$.
Numerical computations presented in this paper have been performed using \textit{Wolfram Mathematica} software.
\begin{figure}[h!]
	\centering
	\hspace{0cm}	(a)\\
	\includegraphics[scale=0.42]{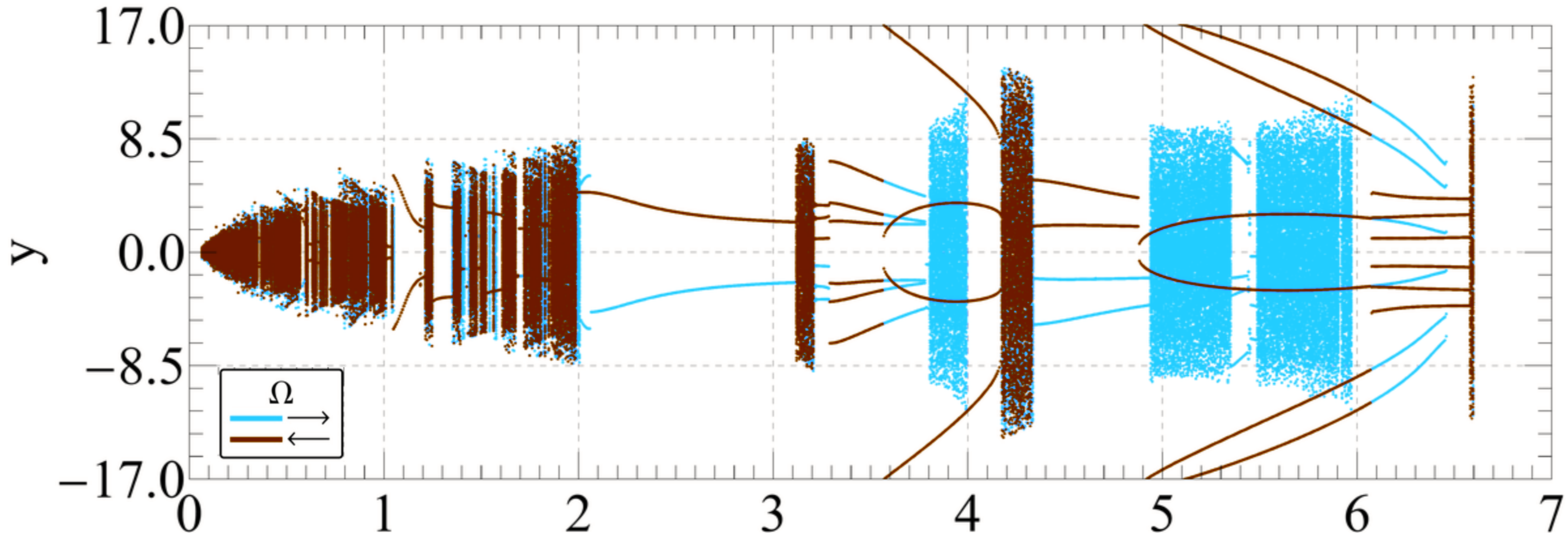}\\
	\hspace{0cm}	(b)\\
	\includegraphics[scale=0.41]{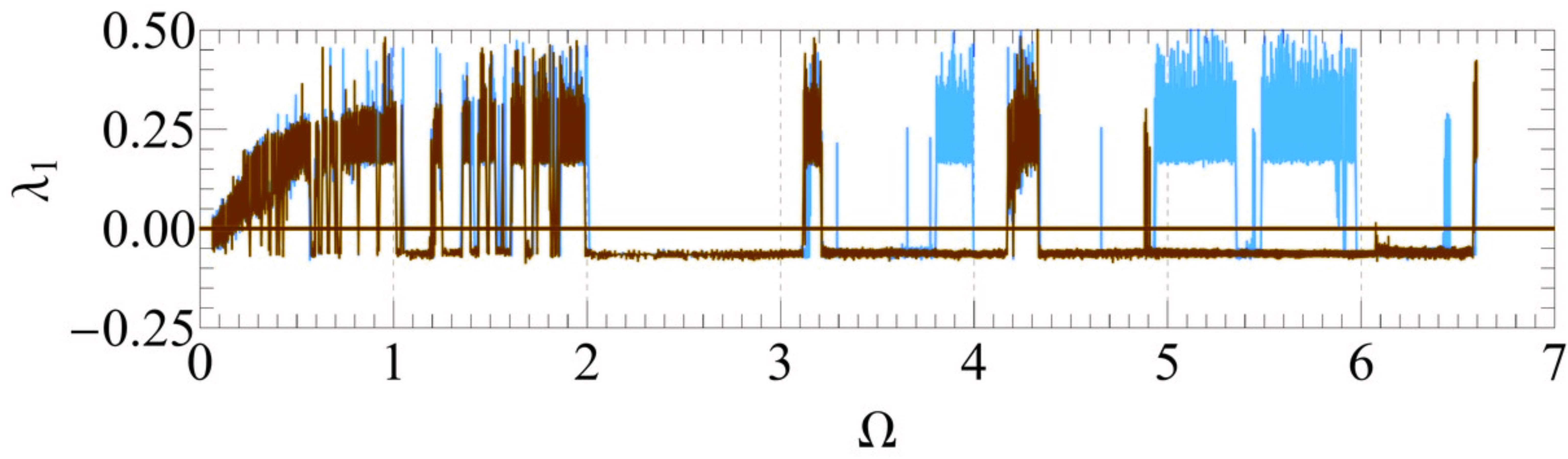}\\
	\hspace{1cm} (c) 	\hspace{4.6cm} (d)\\
	\hspace{0.5cm}	\includegraphics[scale=1]{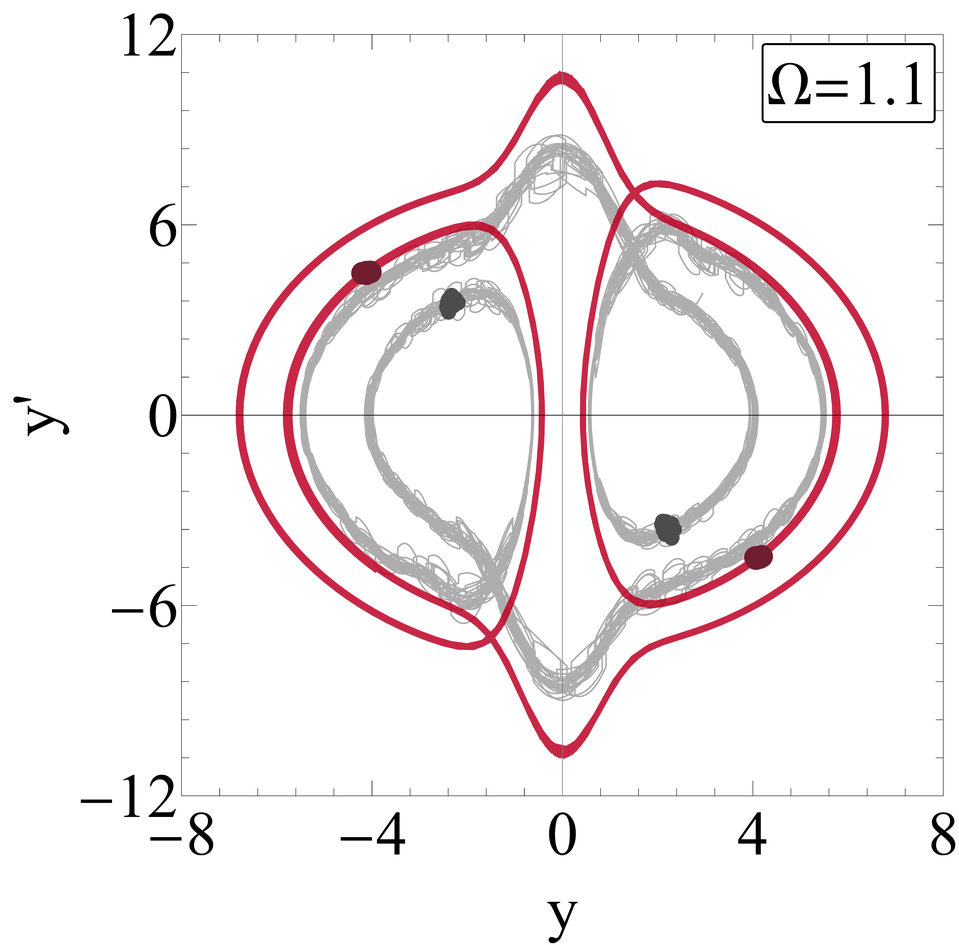}
	\hspace{1cm}	\includegraphics[scale=1]{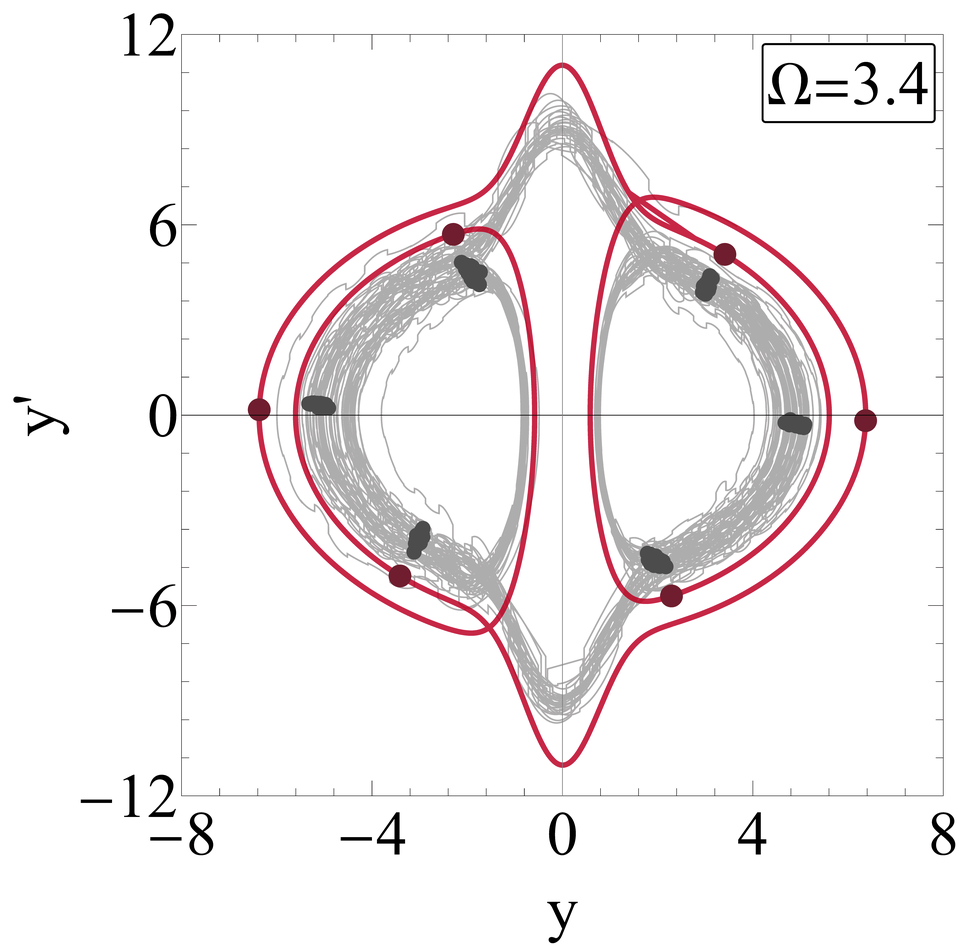}
	\caption{The numerical (a) bifurcation diagram and (b) LLE diagram for the system described by Eq.~(\ref{eq:motion_dimensionless_8}) are presented, with a fixed $A_0=81.82$, showcasing variations in $\Omega$ for both increasing and decreasing values. Simulation (red color) and experimental (gray color) phase planes were obtained for certain values (c,d).}
	\label{fig:bifurcation.png}
\end{figure}
\section{Validation of the model}
Bifurcation analysis was conducted to validate the mathematical model and explore the system's dynamical capabilities. The circular frequency $\Omega$ was chosen as a control parameter, and the bifurcation study involved both increasing and decreasing values, as illustrated in Fig.~\ref{fig:bifurcation.png}.

Bifurcation analysis shows both chaotic and regular dynamics of the systems. Moreover, system exhibits coexisting of periodic attractors, what is visible for example for the range $\Omega=\{2,\, 3.12\}$. Considering the numerous bifurcations of periodic solutions, especially for small values of the control parameter, leading to chaos or qualitatively different solutions, it can be concluded that the system exhibits rich dynamics.
\textcolor{red}{The regular and chaotic dynamics of the system have also been studied using the largest Lyapunove exponent (LLE), see Fig.~\ref{fig:bifurcation.png}b. Positive values of LLE indicate a chaotic regime, while negative ones indicate periodic oscillation.}
For two selected values of the control parameter, phase trajectories were determined and compared with experimental data. Phase trajectories of selected periodic solutions are depicted in Fig.~\ref{fig:bifurcation.png}c,d. A comparison between numerically calculated and experimentally recorded phase trajectories reveals a relatively high agreement sufficient to consider the proposed mathematical model as efficiently reflecting the real system. The difference between the simulation and experimental trajectories may be the result of design inaccuracies in the experimental setup such as clearances in the bearings causing small changes in the distance between the pendulum magnet and the coil during oscillations. The presented nonlinear system is sensitive to changes in the magnetic interaction.
\begin{figure}[b!]
	\hspace{3cm} (a) 	\hspace{5cm} (b)\\
	\hspace{0.5cm}	\includegraphics[scale=1]{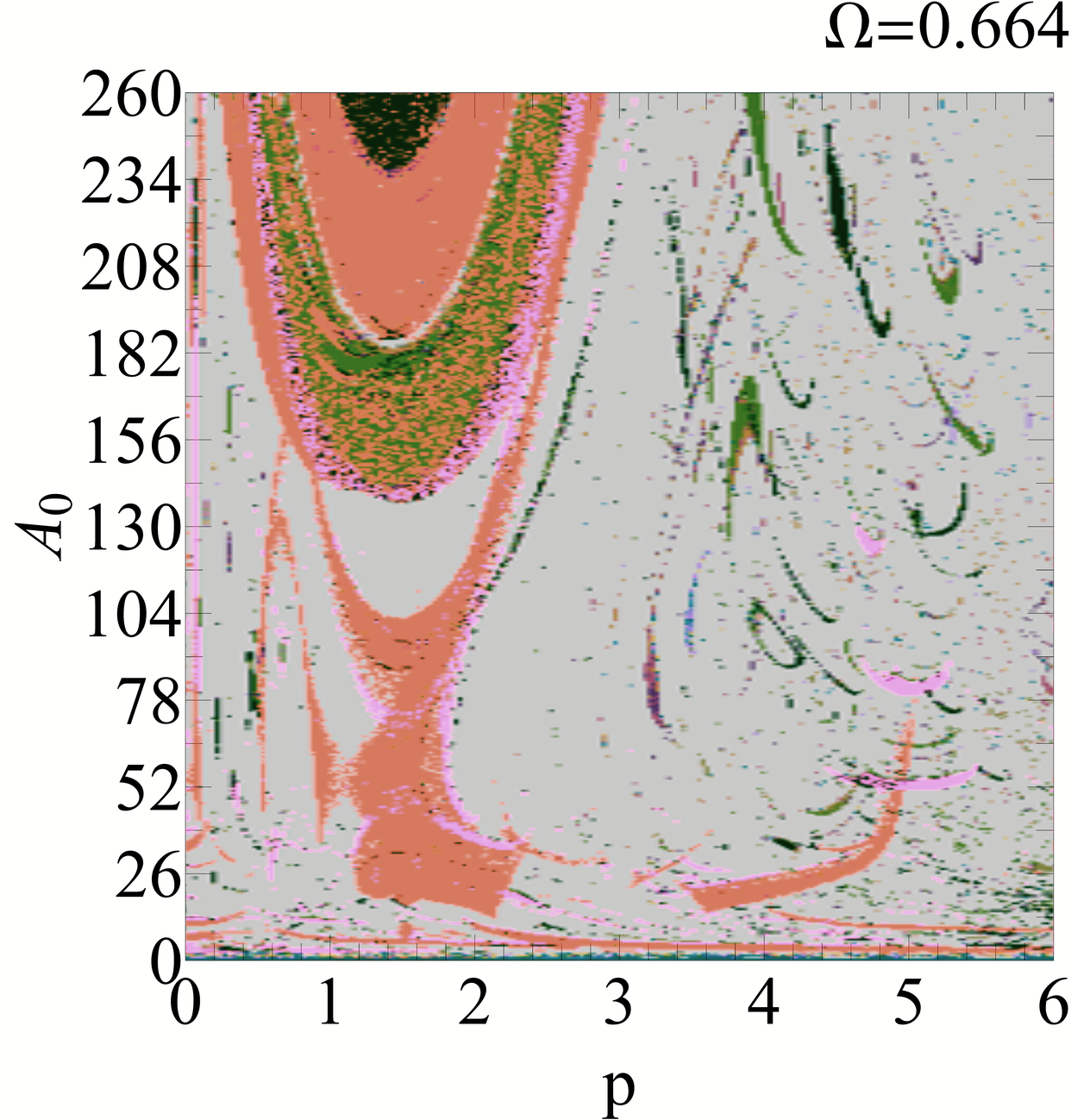}
	\hspace{0.5cm}	\includegraphics[scale=1]{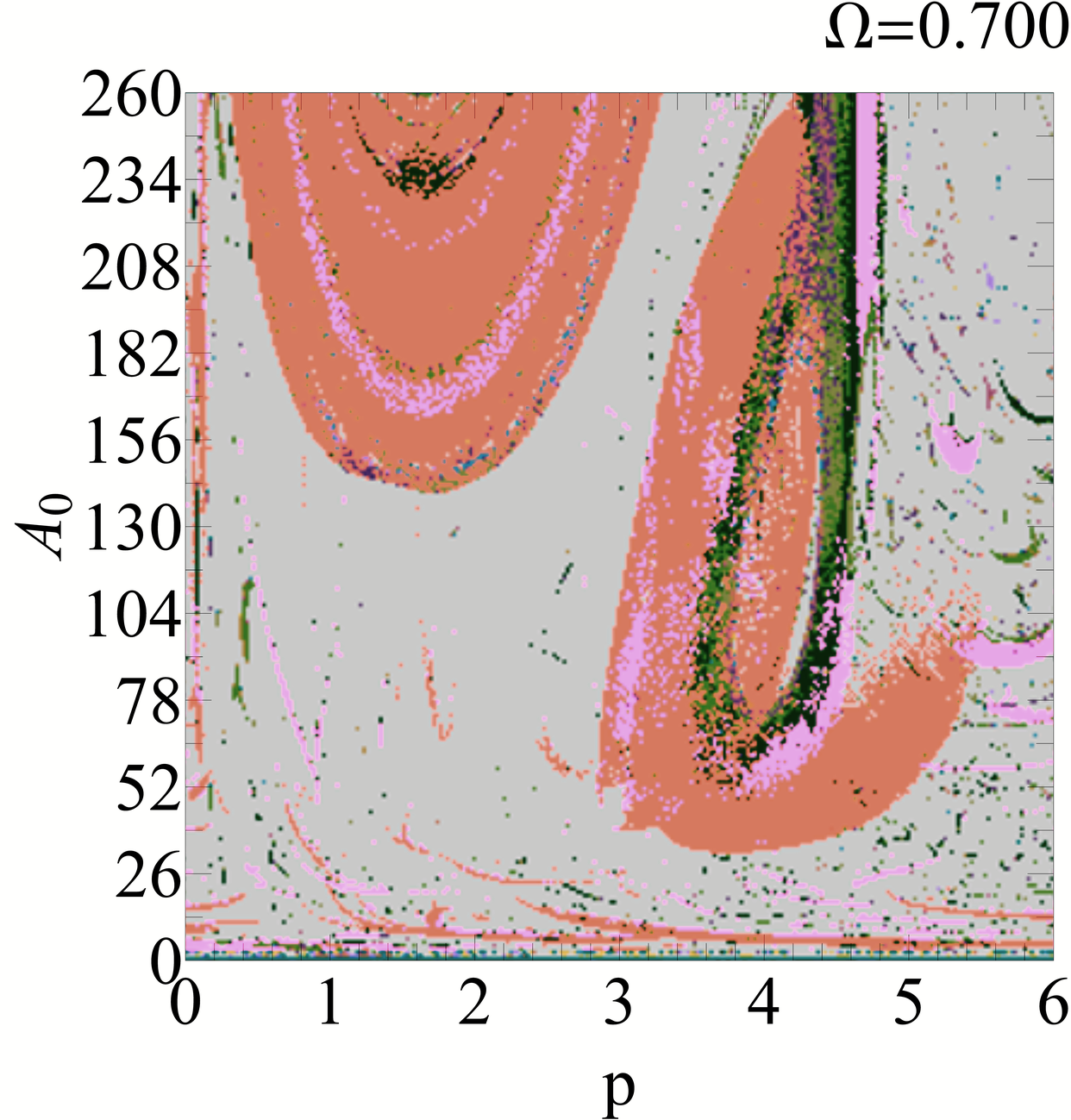}\\
	\hspace{1cm}	\includegraphics[scale=0.13]{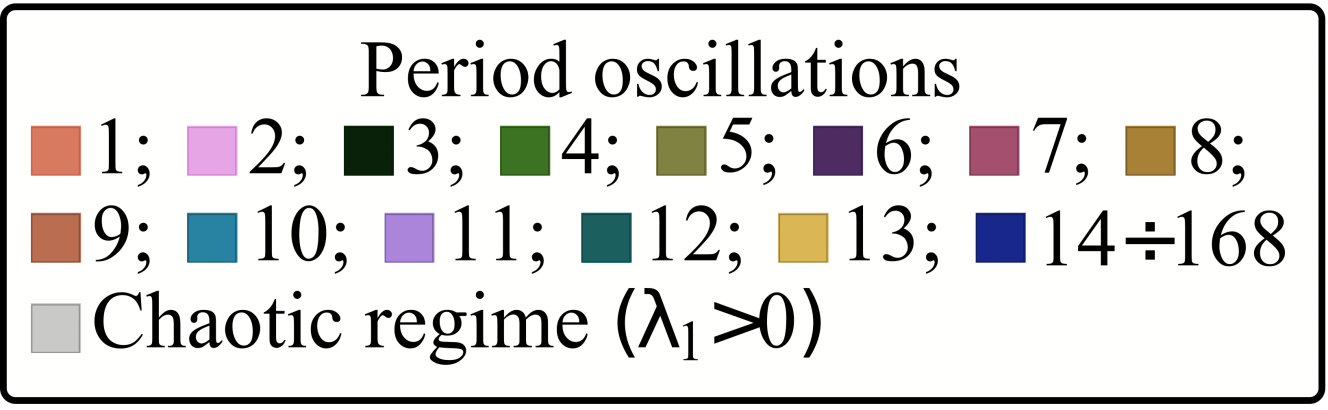}
	\caption{Charts illustrating the dynamical regimes of the system (\ref{eq:motion_control_phase_10}) under excitation with position-dependent and two different frequencies: (a) $\Omega=0.664$, (b) $\Omega=0.700$.}
	\label{fig:phase_control_charts}
\end{figure}
\section{Magnetic pendulum under excitation with position-dependent phase}
\begin{figure}[h!]
	\centering
	\hspace{2cm} (a) 	\hspace{4cm} (b)\\
	\includegraphics[scale=1]{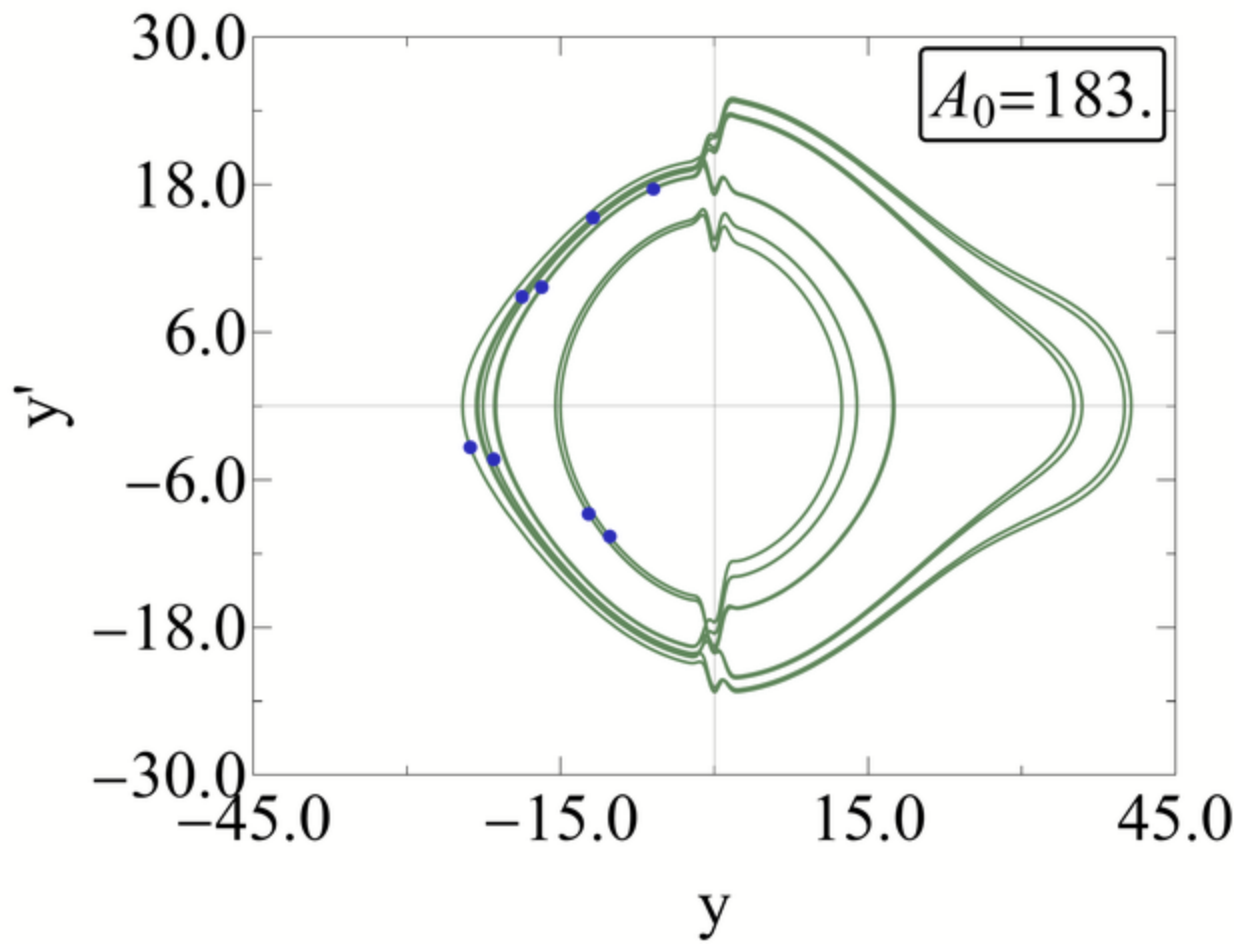}
	\includegraphics[scale=1]{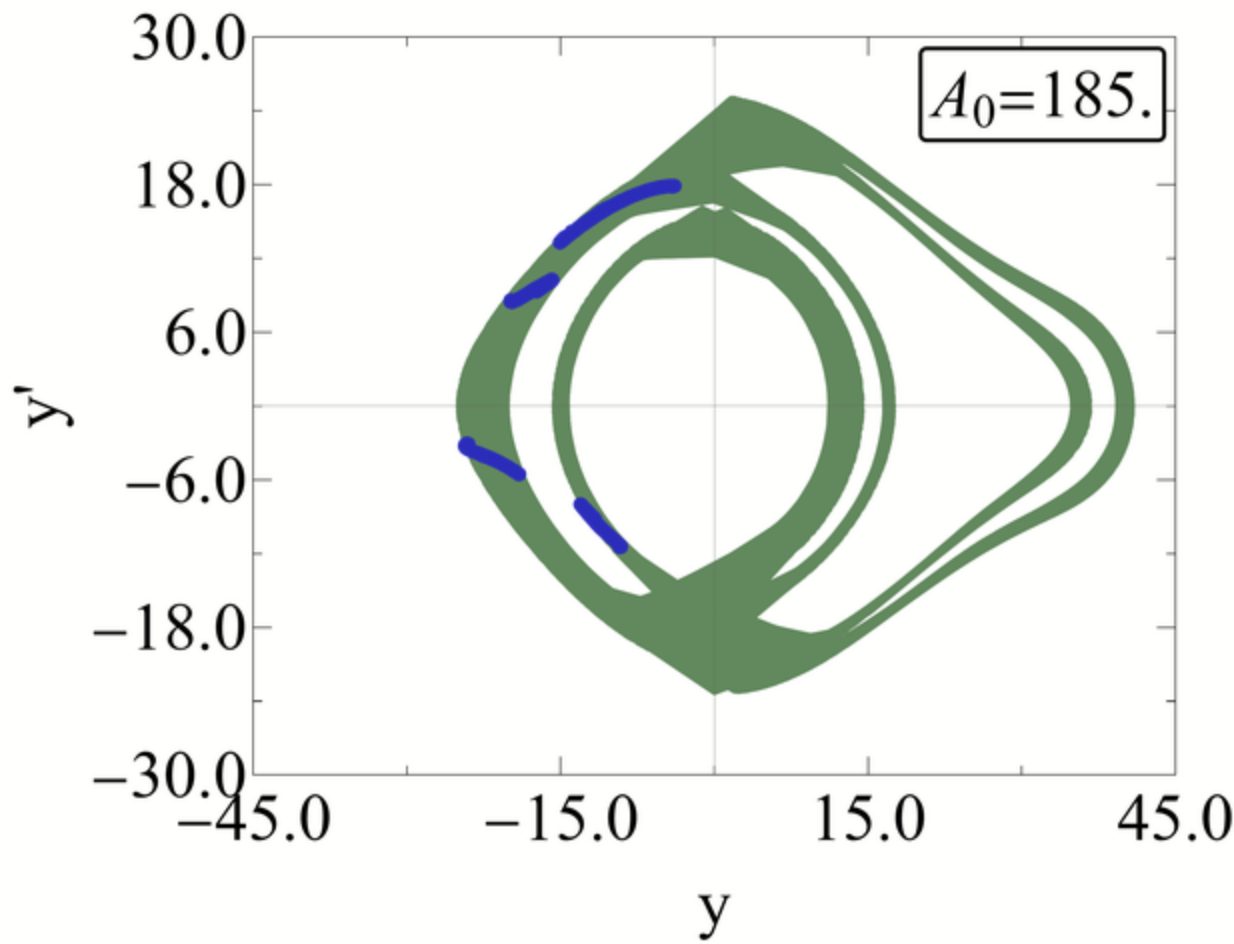}\\
	\hspace{2cm} (c) 	\hspace{4cm} (d)\\
	\includegraphics[scale=1]{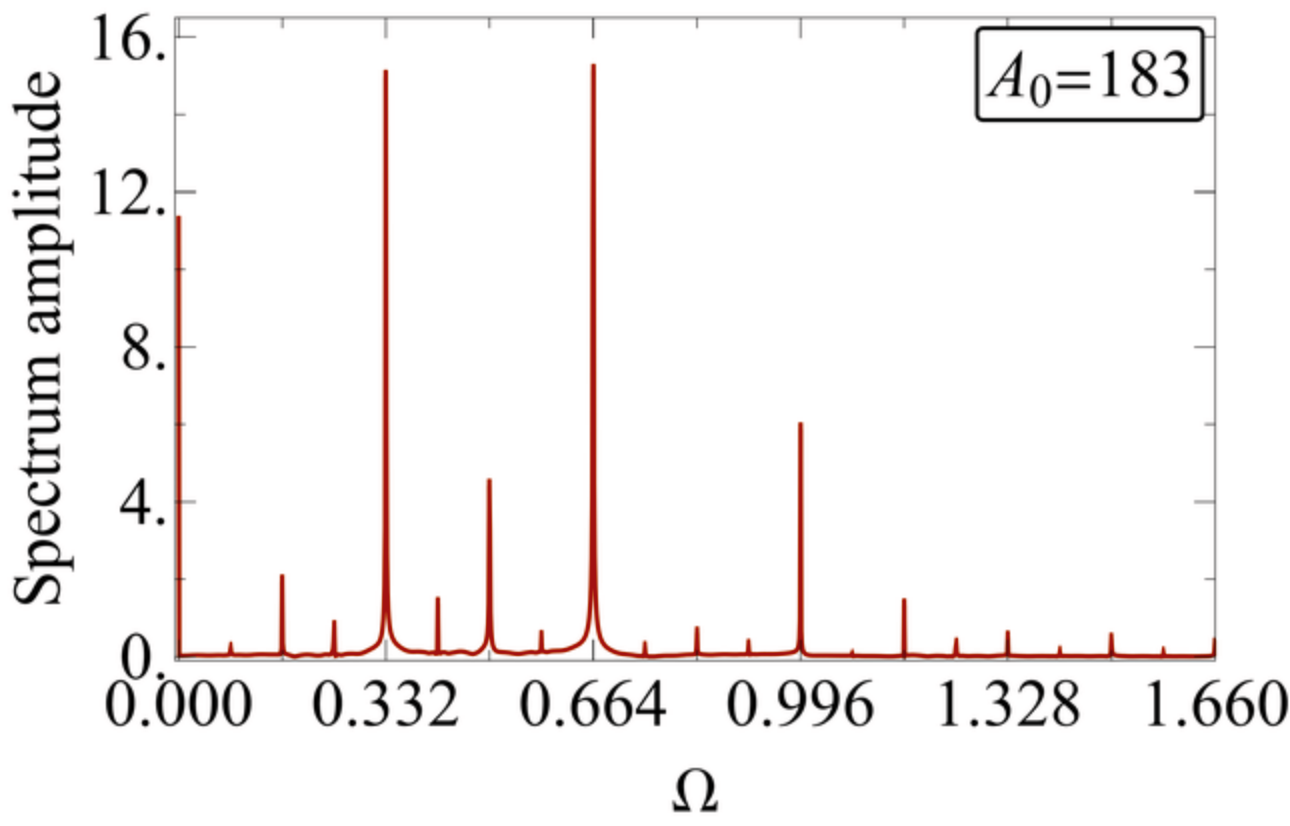}
	\includegraphics[scale=1]{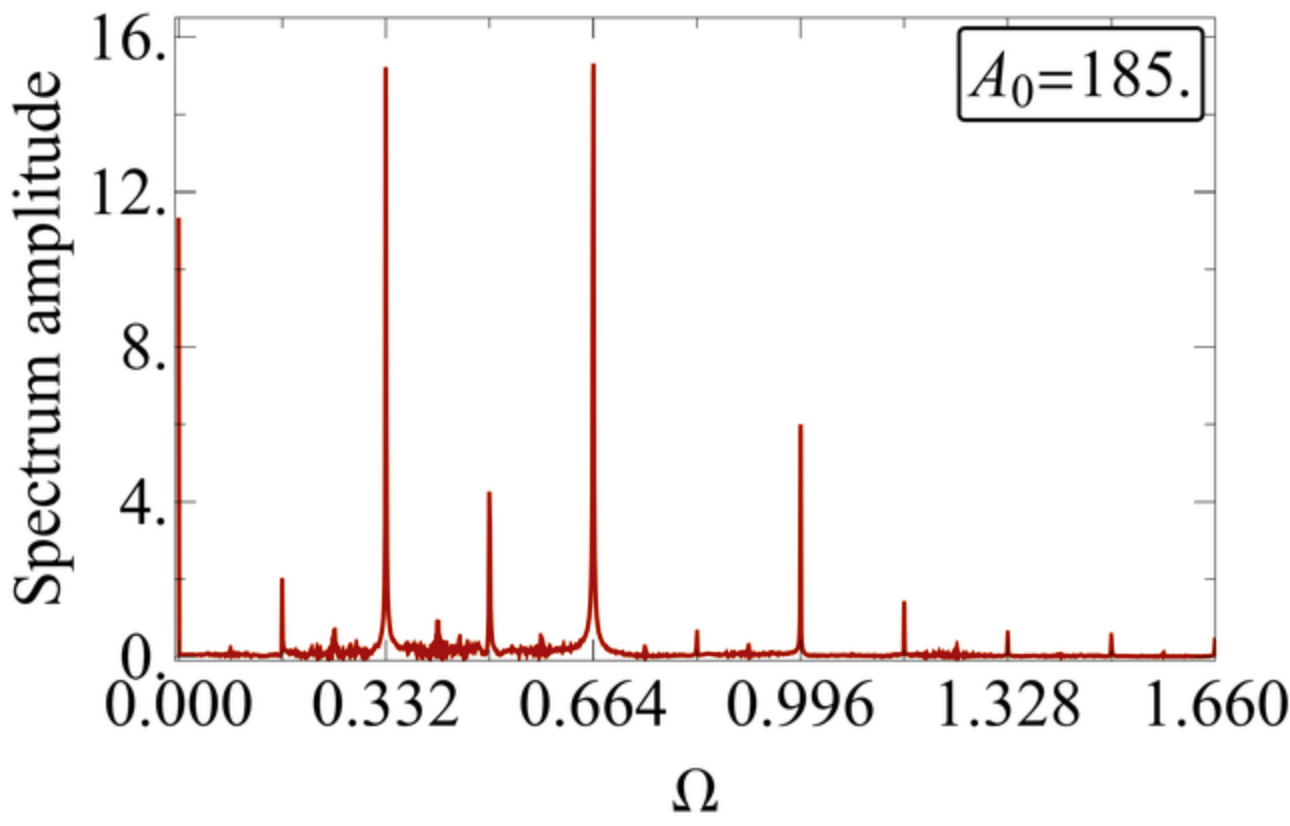}
	\caption{\textcolor{red}{Phase plots and Poincar\'e section (a, b) obtained for periodic and chaotic oscillations, and their frequency spectrums (c, d).}}
	\label{fig:FFT}
\end{figure}

In this section, we present results for the system governed by Eq.~(\ref{eq:motion_dimensionless_8}) with the phase shift $\phi_0$ dependent on the dynamical variable $y$ following a linear function $\phi_0(y) = py$, where $p$ is a constant coefficient. Consequently, the governing dimensionless equation of motion is as follows:
\begin{equation}
	y''+\beta y'+\alpha y+\gamma\sin{\left(\frac{1}{\gamma}y\right)}+\left[\delta+\zeta\exp\left({\nu {y'}^2}\right)\right]\tanh{(\sigma y')}=A_0\exp{(-y^2)}y\cdot\sin{(\Omega x+py)}.
	\label{eq:motion_control_phase_10}	
\end{equation}
Figure \ref{fig:phase_control_charts} displays charts of the dynamical regimes exhibited by the system (\ref{eq:motion_control_phase_10}) on the parameter plane $(p, A_0)$ for the selected excitation frequency $\Omega={0.664,0.700}$. Numerical integrations were conducted with fixed initial conditions $(y,y')=(0.001,0)$ and a time equal to 5000 excitation periods.
\begin{figure}[b!]
	\centering
	\includegraphics[scale=1]{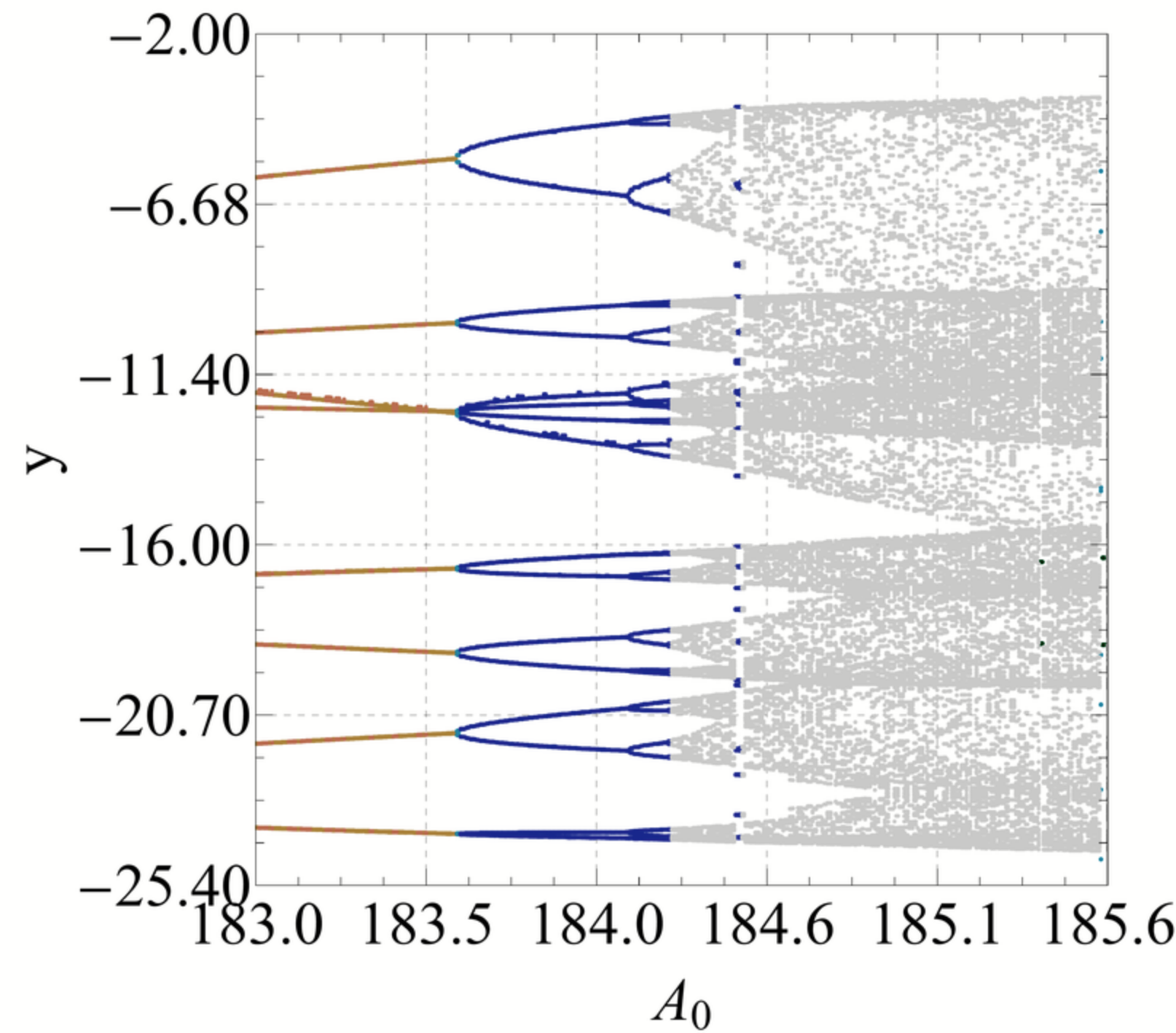}
	\caption{Bifurcation diagram illustrating the route to chaos through a period-doubling sequence.}
	\label{fig:routtochaos}
\end{figure}
Various colors on the charts represent different periodic solutions obtained for the investigated system, and the color legend is provided below the charts. The domain of chaotic oscillations is indicated by the grey color, determined by the positive value of the largest Lyapunov exponent $\lambda_1$. Analyzing the structure of the obtained charts reveals the rich multistability of the system and period-doubling bifurcations.
Various periodic attractors may emerge for the same initial conditions but with slightly different parameters $(p, A_0)$.
As depicted in Fig.~\ref{fig:phase_control_charts}a,b, minor alterations in the excitation frequency $\Omega$ do not lead to significant qualitative changes in the structure of the parameter planes $(p,A_0)$. Changing the frequency mainly affects the bifurcation values of parameters $p$ and $A_0$.
The dynamical charts presented here exhibit a topology similar to those obtained for a classical harmonic oscillator subjected to external forcing with a controlled phase \cite{Krylosova2020}. In both cases, the topology of periodic solution regions resembles elliptical orbits.
To better illustrate the rout to chaos, we created the bifurcation diagram shown in Fig.~\ref{fig:routtochaos}, corresponding to the dynamical chart displayed in Fig.~\ref{fig:phase_control_charts}a, with fixed $p=1.423$ and $A_0\in(183.0,185.6)$. The colors correspond to the periodicity of the solution as before.
Fig.~\ref{fig:FFT}a,b display the phase plots and Poincaré sections for various values of the $A_0$ parameter, illustrating the changes in periodic solutions that, due to doubling, lead to the creation of a chaotic attractor.

The Fourier spectra presented in Fig.~\ref{fig:FFT}c,d have been calculated for the displayed regular period-8 oscillations, as well as chaotic behavior.
The primary peaks in the analyzed FFT spectra correspond to the excitation frequency $\Omega=0.664$ and its multiples. Both period-8 and chaotic motion exhibit the same number of main peaks, with the distinction that the spectrum of the chaotic regime includes noise.
\section{Concluding remarks}
 In conclusion, our investigation into the magnetic pendulum's dynamics under time-varying magnetic excitation with position-dependent phase has revealed a spectrum of intricate behaviors, encompassing both chaotic and regular dynamics. Through a comprehensive blend of numerical simulations and experimental validations, our proposed mathematical model effectively captures the system's richness, highlighting its sensitivity to magnetic interaction variations. The observed chaotic regimes and multistability, including periodic attractors, underscore the complexity of such systems. The agreement between numerical predictions and experimental outcomes reinforces the model's reliability. Our work lays a foundation for further exploration of nonlinear dynamics in magnetic systems and opens avenues for engineering applications harnessing the intricate behaviors uncovered in this study.
\section*{Acknowledgments}
This work was supported by the National Science Center, Poland under the grant \mbox{PRELUDIUM 20} No. 2021/41/N/ST8/01019. For the purpose of Open Access, the authors has applied a CC-BY public copyright licence to any Author Accepted Manuscript
(AAM) version arising from this submission. This is an AAM of an article published by Springer Proceedings in Physics on 10 December 2024, available at: \url{https://doi.org/10.1007/978-3-031-69146-1_25}
\textcolor{red}{
\section*{Appendix A}
	The appendix includes code written in \textit{Wolfram Mathematica} to calculate the bifurcation diagram shown in Fig.~\ref{fig:bifurcation.png}a.}
	
	\begin{adjustwidth}{-5.8cm}{-0.8cm} % Dostosuj marginesy wewnątrz lstlisting
		\begin{lstlisting}[language=Mathematica]
			eqbez[f_]:=y''[x]+$\beta$ y'[x]+$\alpha$ y[x]+$\gamma$ Sin[1/$\gamma$ y[x]]+
			($\delta$+$\lambda$ E$^{\nu y'[x]^2}$) Tanh[$\sigma$ y'[x]]-A0 E$^{-y[x]^2}$ y[x] Sin[$\Omega$(f)x+$\phi$0]==0 
			(*governing eq.*);
			
			$\phi$ = 0.0001; d$\phi$ = 0; (* initial angle and velocity *)
			
			biffurcationwym[ndrop_, nplot_, parChange_, init_] := (
			TB[f_] := 1/f;
			eventB[f_] := WhenEvent[Mod[t, TB[f]] == 0, Sow[{$\phi$[t], 
							$\phi$'[t]}]];
			funk[arg_, par_] := {par, #} & /@ Drop[Flatten[
			Reap[NDSolve[{eqbez[par], $\phi$[0]==arg[[-1, 2, 1]], $\phi$'[0]==arg[[-1, 2, 2]], 
				eventB[Echo[par]]}, {}, {t, 0, TB[par]*(ndrop + nplot)},Method ->{"StiffnessSwitching","DiscontinuityProcessing"->False},MaxSteps->$\infty$]]][[2]], 1], ndrop];
			
			biffData = Drop[Flatten[FoldList[funk, init, parChange], 1], 1];
			(* Save to file *)
			Export[ToString[NotebookDirectory[]] <>	"bif_" <> ToString[parChange[[1]]] <> "_" <> 
			ToString[parChange[[-1]]] <> "Hz" <> ToString[ndrop] <> "_" <> ToString[nplot] <> "_" <> ".m", biffData]
			)
			
			init = {{0, {$\phi$, d$\phi$}}}
			parChange =(* Reverse@ *) Table[i, {i, 0.1, 10, 0.01}];
			biffurcationwym[10000, 200, parChange, init];
		\end{lstlisting}
	\end{adjustwidth}
% ---- Bibliography ----
%

\end{document}